\documentclass[final,twocolumn]{elsarticle}
\journal{Chinese Journal of Physics}
\usepackage{tikz}
\usepackage{pgfplots}
\usepackage{xcolor}
\usepackage{graphicx}
\usepackage{dcolumn}
\usepackage{braket}
\usepackage{bm}
\usepackage{amsfonts}
\usepackage{amsmath}
\usepackage{amssymb}
\usepackage{color,soul}
\usepackage{wasysym}
\usepackage{mathrsfs}
\usepackage{float}
\usepackage{multirow}
\usepackage{mathtools}
\usetikzlibrary{spy}
\usepackage[T1]{fontenc}
\usepackage[utf8]{inputenc}

\usepackage[colorlinks=true,%
            linkcolor=blue,%
            urlcolor=blue,%
            citecolor=blue,%
            filecolor=blue,%
            bookmarksopen=true,%
            pdfauthor={yo},%
            pdftitle={CSSH},%
            pdfsubject={CSSH_Manuscript},%
            pdfpagemode=UseOutlines]{hyperref}

\usepackage{orcidlink}
\newcommand{\orcid}[1]{\textsuperscript{\orcidlink{#1}}}

\begin{document}

\begin{frontmatter}

\title{Transport properties and topological phase transitions for a Creutz-Su-Schrieffer-Heeger ladder}

\author[utfsm,pucv]{K. A. González}
\author[utfsm]{S. Bravo\orcid{0000-0001-5165-7916}}
\author[utfsm]{L. Rosales\corref{cor1}\orcid{0000-0002-9725-5522}}
\ead{luis.rosalesa@usm.cl}
\author[utfsm]{P. A. Orellana\orcid{0000-0001-7688-4111}}

\cortext[cor1]{Corresponding author}
\address[utfsm]{Departamento de Física, Universidad Técnica Federico Santa María, Casilla 110 V, Valparaíso, Chile}
\address[pucv]{Instituto de Física, Pontificia Universidad Católica de Valparaíso, Casilla 4059, Valparaíso 2362804, Chile}

\begin{abstract}
In this work, we investigate the electronic, topological, and transport properties of a Creutz-Su-Schrieffer-Heeger (CSSH) ladder. Using a tight-binding model within the Green's function formalism, we calculate the energy spectrum, local density of states (LDOS), and electronic transmission. We first determine the energy spectrum of the CSSH ladder and analyze the different topological phases present in the system, identifying one trivial phase and three distinct nontrivial regions. We then study electronic transport and show that the transmission reproduces the different topological phases through characteristic transport signatures. Finally, we derive the conditions for the emergence of non-topological flat bands and demonstrate that these bands also provide the necessary conditions for the formation of bound states in the continuum (BICs). Our results establish a direct connection between the topological properties, flat-band formation, and electronic transport in the CSSH ladder. 
\end{abstract}

\begin{keyword}
Creutz-Su-Schrieffer-Heeger ladder \sep topological phase transitions
\sep electronic transport \sep bound states in the continuum
\end{keyword}

\end{frontmatter}

\section{Introduction}

Topological insulators (TIs) are a distinct class of materials that behave as insulators in the bulk while hosting conducting states at their boundaries. Their hallmark feature is robust edge states that can maintain their conductivity even in the presence of perturbations. The Su-Schrieffer-Heeger (SSH) model \cite{PhysRevLett.42.1698} provides a prototypical example of this physics. It describes a one-dimensional (1D) dimerized chain with distinct intercell and intracell hopping amplitudes. When the intercell hopping is greater than the intracell hopping, the model enters a topologically nontrivial phase characterized by edge states, otherwise, it remains topologically trivial.
The SSH model has been extensively investigated across various domains, including optical systems \cite{Khanikaev2013,Hotte-Kilburn_2024}, ultracold atoms \cite{Xie2019,PhysRevA.97.023618}, superconducting systems \cite{PhysRevB.106.054511}, and as platform for engineering quantum state transfer \cite{PhysRevA.98.012331,PhysRevA.102.022404}.

A generalization of the SSH model is the CSSH ladder, a term introduced to describe the combination of the SSH and Creutz ladder structures. The Creutz ladder \cite{PhysRevLett.83.2636} is a BDI topological insulator that can host flat bands either through Aharonov-Bohm (AB) caging in the presence of a magnetic flux or through interference effects under specific parameter conditions, with or without magnetic flux \cite{https://doi.org/10.1002/qute.201900105}. Under these conditions, the states forming each flat band become dark states and behave as bound states in the continuum (BICs) \cite{1929PhyZ...30..467V}. When a weak symmetry-breaking perturbation is introduced, these BICs can couple to the surrounding continuum states, giving rise to quasi-BICs. The CSSH ladder has been investigated in several contexts, including the emergence and control of tunable zero-energy modes \cite{Zurita2021tunablezeromodes}, the effects of two-dimensional (2D) model of a weak topological insulator with an $N$
stacked Su-Schrieffer-Heeger chains \cite{PhysRevB.108.104101}, and the topological insulator behavior in a ladder model with particle-hole and reflection symmetries \cite{Hetenyi_2018}.

\begin{figure}[h!]
\includegraphics[width=\columnwidth]{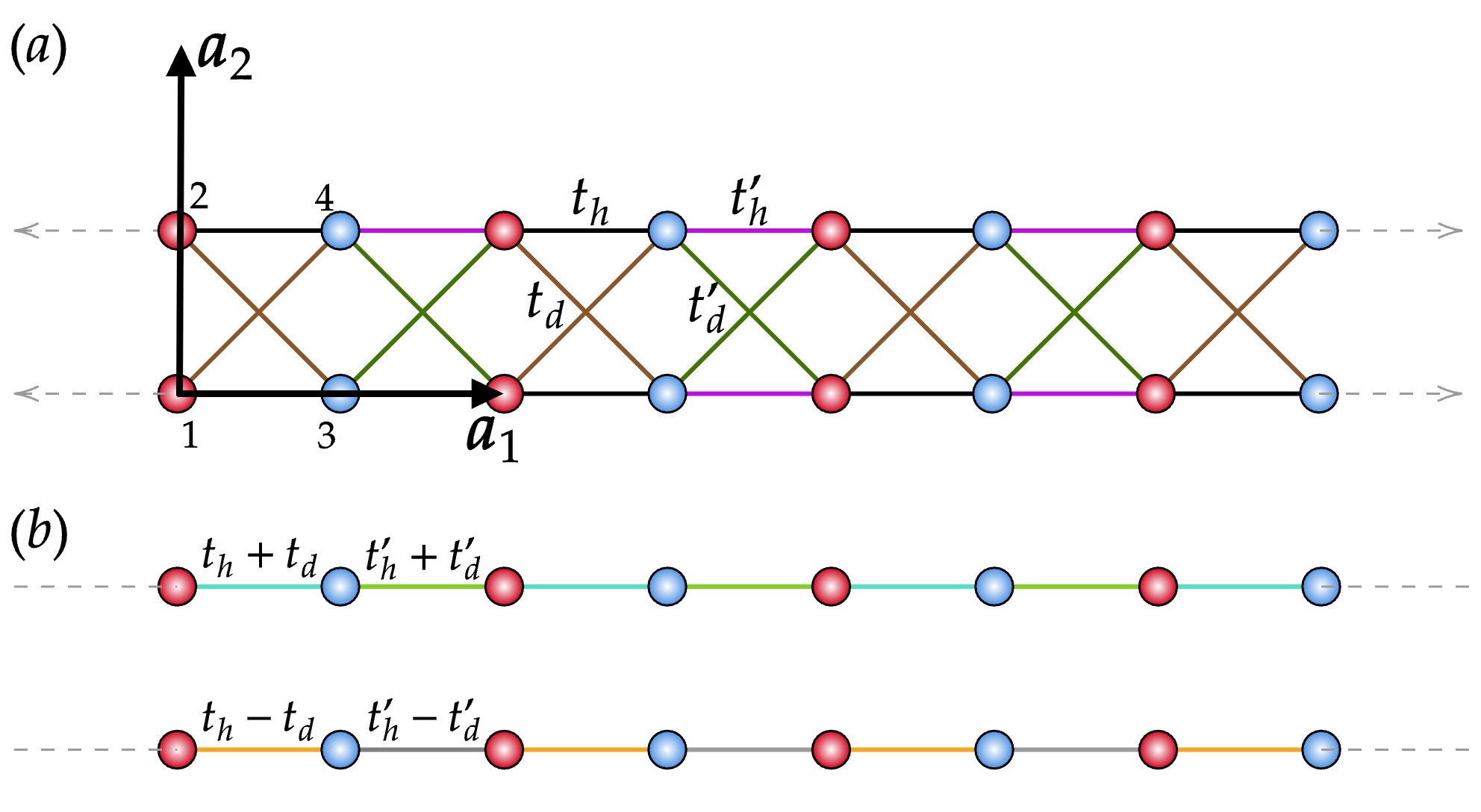}
\caption{(a) Real space structure of the Creutz ladder lattice with hopping amplitudes
labeling. (b) Representation in terms of effective SSH chains.}
\label{fig1}
\end{figure}

In this context, we investigate the electronic, topological, and transport properties of a Creutz-Su-Schrieffer-Heeger (CSSH) ladder in the absence of a magnetic field, with special attention to the topological edge states associated with the SSH structure and the non-topological flat bands and bound states in the continuum (BICs) originating from the Creutz lattice. We first analyze the bulk properties of the system by calculating its energy spectrum and topological phases through the diagonalization of the Hamiltonian into two independent effective channels, each equivalent to an SSH chain with renormalized hopping parameters. This approach allows us to identify the trivial and nontrivial topological phases, determine the conditions for the formation of flat bands, and establish the different localization mechanisms supported by the system. We then study the electronic transport in finite CSSH ladders within the Green's function formalism, showing that the transmission correctly reproduces the topological phase diagram and provides clear transport signatures of the corresponding edge states. Finally, we derive the conditions for the emergence of non-topological flat bands and demonstrate that they also provide the necessary conditions for the formation of BICs, which appear as Dirac $\delta$-function peaks in the local density of states. Our results establish a direct connection between the topological phases, flat-band formation, and electronic transport in the CSSH ladder, showing that topological edge states and BICs originate from different localization mechanisms while together determining the electronic response of the system.

\section{Model, bulk spectrum and topology}\label{model}

FIG. \ref{fig1}(a) shows the geometry of the CSSH ladder in real space, where the structure is built from a square unit cell defined by the lattice vectors $\boldsymbol{a}_{1}=(a,0)$ and $\boldsymbol{a}_{2}=(0,a)$, with $a$ denoting the lattice constant. Throughout the following analysis, we set $a=1$ and take the periodic direction along $\boldsymbol{a}_{1}$. Each unit cell contains four sites, with one state per site, denoted by $|\psi_i\rangle$, where the site label satisfies $i\in\{1,2,3,4\}$. All states have the same internal structure. The positions of the sites within the unit cell are $\boldsymbol{r}_{1}=(0,0)$, $\boldsymbol{r}_{2}=a(0,1/2)$, $\boldsymbol{r}_{3}=a(1/2,0)$, and $\boldsymbol{r}_{4}=a(1/2,1/2)$.

In real space, each unit cell in the ladder is labeled by an integer index $n$, so any state within the ladder is identified by the pair $(n,i)$. In this notation, $c_{n,i}^{\dagger}$ ($c_{n,i}$) denotes an operator that creates (annihilates) a particle in state $i$ at unit cell $n$. Using these operators, the Hamiltonian of the system in real space can be written as follows.
\begin{equation}
H=H_{t_{h}} +H_{t'_{h}} +H_{t_{d}} +H_{t'_{d}},
\end{equation}

Each of these terms represents a distinct type of hopping amplitude. Four horizontal hoppings are included: two intracell ($t_h$) and two intercell ($t'_h$). Diagonal hoppings are arranged similarly, comprising two intracell ($t_d$) and two intercell ($t'_d$) couplings. Since all states are considered equivalent, all onsite energies are set to zero. The explicit form of each interaction term is provided below:
\begin{equation}
\begin{aligned}
H_{t_{h}} & =t_{h}\sum _{n} c_{n,1}^{\dagger } c_{n,3} +c_{n,2}^{\dagger } c_{n,4} \ +h.c\ ,\ \ \\
H_{t'_{h}} & =t'_{h}\sum _{n} c_{n,3}^{\dagger } c_{n+1,1} +c_{n,4}^{\dagger } c_{n+1,2} \ +h.c\ ,\ \\
H_{t_{d}} & =t_{d}\sum _{n} c_{n,1}^{\dagger } c_{n,4} +c_{n,2}^{\dagger } c_{n,3} \ +h.c\ ,\\
H_{t'_{d}} & =t'_{d}\sum _{n} c_{n,4}^{\dagger } c_{n+1,1} +c_{n,3}^{\dagger } c_{n+1,2} \ +h.c\ .
\end{aligned}
\end{equation}

Applying the Fourier transformation to $H$ in reciprocal $k$-space, which is one-dimensional for this system, yields a Bloch Hamiltonian $H(k)$ in block matrix form:
\begin{equation}
H( k) =\begin{pmatrix}
0_{2} & Q( k)\\
Q^{\dagger }( k) & 0_{2}
\end{pmatrix} ,
\end{equation}
where $0_{2}$ denotes the $2\times2$ zero matrix, and the $2\times2$ matrix $Q(k)$ is given by:

\begin{equation}
Q( k) =\begin{pmatrix}
t_{h}+t'_{h} e^{-ik} & t_{d} +t'_{d} e^{-ik}\\
t_{d} +t'_{d} e^{-ik} & t_{h}+t'_{h} e^{-ik}
\end{pmatrix} .
\end{equation}

From the block structure of $H(k)$, it is straightforward to see that the system possesses chiral symmetry. This symmetry is represented by an operator $\Gamma$ satisfying $\Gamma H(k)\Gamma^{-1}=-H(k)$ and $\Gamma^2=1$. Thus,

\begin{equation}
\Gamma =\begin{pmatrix}
I_2  & 0\\
0 & -I_2 
\end{pmatrix} ,
\end{equation}
where $I_2$ denotes the $2 \times 2$ identity matrix. 

The system preserves time-reversal symmetry $\mathcal{T}$. Since we consider a spinless system, where $\mathcal{T}^2=1$, the time-reversal operator is simply given by $\mathcal{T}=\mathcal{K}$, with $\mathcal{K}$ denoting the complex conjugation operator. The combination of time-reversal and chiral symmetries results in an emergent particle-hole symmetry $\mathcal{P} = \Gamma \mathcal{T}$, with $\mathcal{P}^2 = 1$. According to the Altland–Zirnbauer (AZ) classification, the Creutz ladder as defined above belongs to the BDI class \cite{Altland1997}. Within this topological framework for a one-dimensional BDI system, stable nontrivial topology can exist, and the different phases are distinguished by a $\mathbb{Z}$-valued bulk invariant, denoted by $\nu$. Although the system possesses spatial symmetries, these are not essential for the present topological discussion.

For a periodic one-dimensional system in the BDI class, $\nu$ can be computed in momentum space directly from the nonzero block of $H(k)$ in the form:
\begin{equation}
\nu =\frac{1}{2\pi i}\int _{-\pi}^{\pi} dk\partial _{k} ln(detQ(k)).
\end{equation}

The determinant of $Q$ can be obtained in closed form in this model such that:

\begin{equation}
detQ( k) =F_{+}(k)F_{-}( k)
\end{equation}
and 
\begin{equation}
F_{\pm}(k)=(t_{h} \pm t_{d}) +(t'_{h} \pm t'_{d})e^{-ik} .
\end{equation}

This factorization of $detQ$ implies that the invariant can be expressed as the sum
of two contributions, namely:
\begin{equation}
\nu =\frac{1}{2\pi i}\int _{-\pi }^{\pi } dk\partial _{k} ln( F_{+})+
\frac{1}{2\pi i}\int _{-\pi }^{\pi } dk\partial _{k} ln( F_{-}),
\end{equation}
which implies that: 
\begin{equation}
\nu =\nu _{+} +\nu _{-}.
\label{eq:windingnumbersum}
\end{equation}

This suggests that the CSSH ladder can be represented as two effective SSH chains, one characterized by the form factor $F_{+}$ and the other by $F_{-}$. This configuration is illustrated in FIG. \ref{fig1}(b). In terms of topology, the chains contribute additively, as they correspond to decoupled chiral channels within the system. Consequently, $\nu$ is expressed as a combination of two SSH invariants. The SSH invariant is known to assume only two values, $\nu_{\pm} \in \{0,1\}$. Therefore, $\nu \in \{0,1,2\}$, indicating that only a subset of the full range of the BDI invariant is accessible in this system. This is expected since topology in this class is determined by the number of independent chiral channels the system can support, which is limited to two in this case. Higher invariant values may be realized by constructing a multi-leg Creutz ladder within the BDI class \cite{Lahiri2023, Nandy2023}.

Further insights can be extracted from the analysis of the functions $F_{\pm }$, which
have the generic form given by:
\begin{equation}
F_{\pm }( k) =A_{\pm } +B_{\pm } e^{-ik},
\end{equation}

where $A_{\pm} = t_{h} \pm t_{d}$ and $B_{\pm} = t'_{h} \pm t'_{d}$. When these functions are defined in the complex plane, as $k$ traverses a complete cycle in reciprocal space, $F_{\pm}$ traces a circle centered at $A_{\pm}$ with radius $|B_{\pm}|$. This geometric representation directly illustrates the nontrivial topological behavior as a winding in the complex plane.

In addition to the winding analysis, the most general form of the state space for the Creutz model, characterized by the five-dimensional parameter space $(k, t_{h}, t'_{h}, t_{d}, t'_{d})$, can be examined alongside the constraints imposed by $F_{\pm}$. Identifying regions in parameter space where the bulk spectral gap closes provides the foundational information for outlining the topological phases in this context. These regions are critical, as they delineate the coexistence of different topological phases and indicate parameter variations that may facilitate transitions between phases.
The analysis of the invariant $\nu$ indicates that topological transitions and the associated spectral gap closings occur when $\det Q(k) = 0$. This condition corresponds to $F_{+}(k) = 0$, $F_{-}(k) = 0$, or both, which is equivalent to a closing of the energy spectrum $E(k) = 0$. The following section examines all possible cases.

Let us begin with the condition $F_{+}(k)=0$. Separating it into its real and imaginary parts gives:
\begin{equation}
\begin{aligned}
A_{+} + B_{+}\cos(k) &=0,\\
B_{+}\sin(k) &=0 .
\end{aligned}
\label{critical_zones_eq}
\end{equation}
We first focus on the case in which both $A_{+}$ and $B_{+}$ are nonzero, leaving the analysis of the cases where either coefficient vanishes to a later section, where the flat-band regime is discussed. Under this assumption, the second equation requires $k=0$ or $k=\pi$, so gap closings can occur only at these two high-symmetry points. The parameter space is therefore reduced to a four-dimensional subspace at fixed $k$, within which we analyze the geometry of the gap-closing regions.

Inserting these values of $k$ into the first equation yields two independent conditions for the hopping parameters, given by:
\begin{equation}
t_{h} +t_{d} =\pm (t'_{h}+t'_{d}).
\end{equation}

Each case imposes a single constraint on the four independent parameters and therefore defines a codimension-1 subspace in the $(t_h,t'h,t_d,t'd)$ parameter space. These two codimension-1 regions are complemented by two additional regions that arise from the condition $F_{-}=0$ with $F_{+}\neq 0$, for which gap closings also occur when:
\begin{equation}
t_{h}-t_{d} =\pm (t'_{h}-t'_{d}).
\end{equation}

\begin{figure}[ht]
    \includegraphics[width=1.\linewidth]{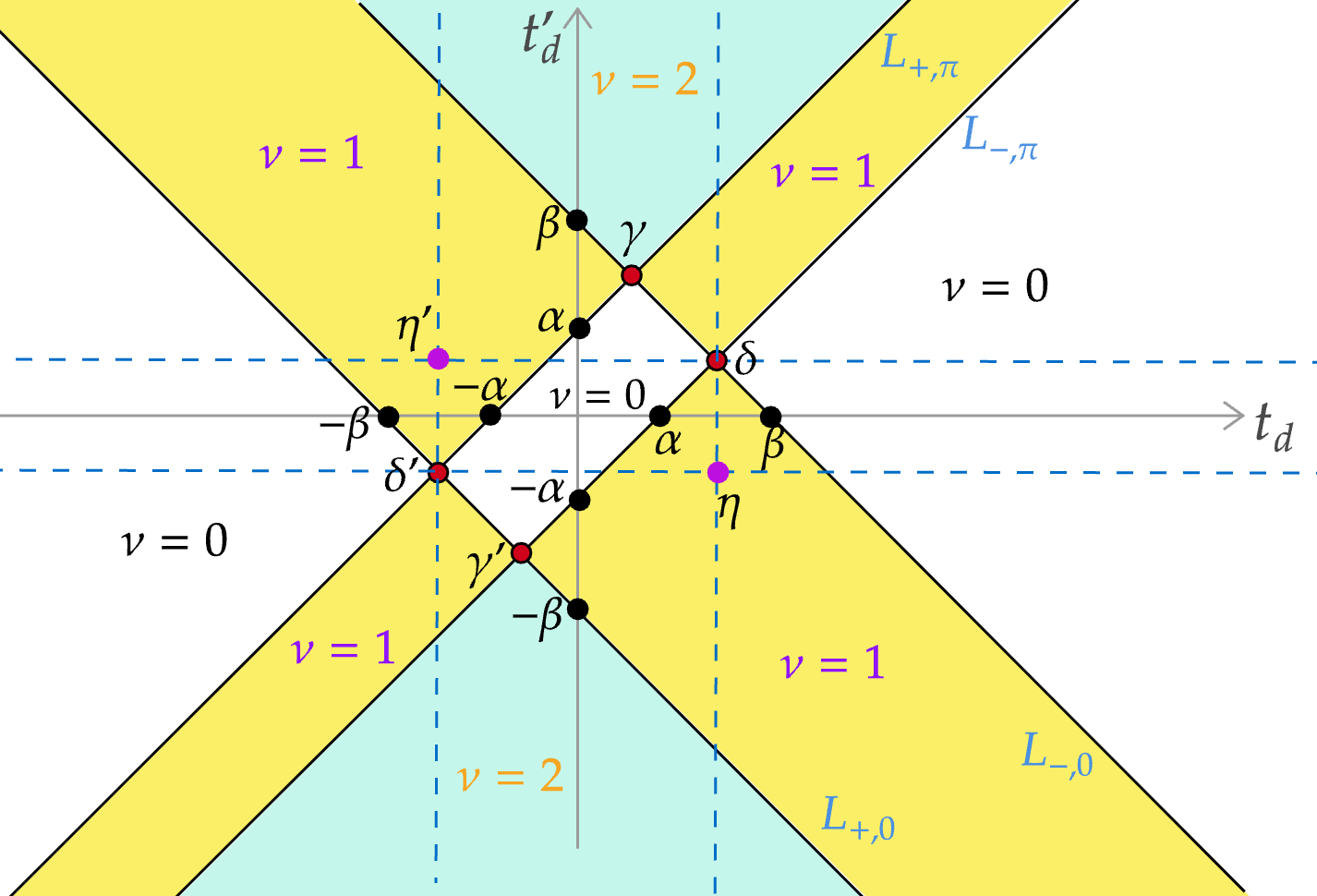} 
\caption{Two-dimensional projection of the four-dimensional bulk phase diagram for the topological phases of the CSSH ladder. Solid lines represent the gap-closing 
regions in the spectrum, while dashed lines indicate the flat-band formation conditions. The labels $\alpha$ and $\beta$ denote points at which, in addition to lying on a topological phase-transition line, one of the intercell hopping parameters vanishes. The labels $\eta$ and $\eta'$ correspond to codimension-2 regions where two flat bands become degenerate. The points $(\delta,\delta',\gamma,\gamma')$ indicate parameter values at which simultaneous gap closings occur in momentum space. In addition, $(\delta,\delta')$ also 
correspond to points where two flat bands become degenerate.} \label{bulk_Phase_diag}
\end{figure}

Thus, four codimension-1 regions can be identified, across which topological phase transitions occur. To visualize these transition regions, we project the corresponding conditions onto two-dimensional subspaces. A convenient choice is to fix the horizontal hopping amplitudes $t_h$ and $t'_h$ and analyze the resulting $(t_d,t'_d)$-plane. Under this projection, the critical regions appear as lines in the two-dimensional parameter space. FIG. \ref{bulk_Phase_diag} shows the projected phase diagram for $|t_h|>t'_h$. In particular, we set $t'_h=1$, so that all other hopping amplitudes are measured in units of $t'_h$ hereafter.

The gap-closing regions discussed above are represented by the solid lines in FIG. \ref{bulk_Phase_diag}, collectively denoted by $L_{j,k}$. Here, $j\in\{+,-\}$ specifies which of the functions $F_{\pm}$ vanishes, $k\in\{0,\pi\}$ indicates the momentum at which the gap closing occurs, and the labels $\alpha$ and $\beta$ are defined in the figure caption. This phase diagram also reveals additional structure within the projected parameter space. In particular, we focus on the natural extension of the codimension-1 regions, which correspond to the four intersections between pairs of solid lines (temporarily disregarding the dashed lines of this figure). These intersections identify parameter values at which two gap closings occur simultaneously, one at $k=0$ and the other at $k=\pi$. This requires imposing one of the following sets of conditions:
\begin{equation*}
\begin{array}{ c c c }
\gamma : & \ F_{+}(\pi) =0\  & F_{-}(0)=0.\\
\gamma ': & F_{+}(0) =0\  & F_{-}(\pi)=0.\\
\delta : & F_{-}(0) =0\  & F_{-}(\pi)=0.\\
\delta ': & F_{+}(0) =0\  & \ F_{+}(\pi)=0.\ 
\end{array}
\end{equation*}

\begin{figure*}[ht]
    \includegraphics[width=1.\linewidth]{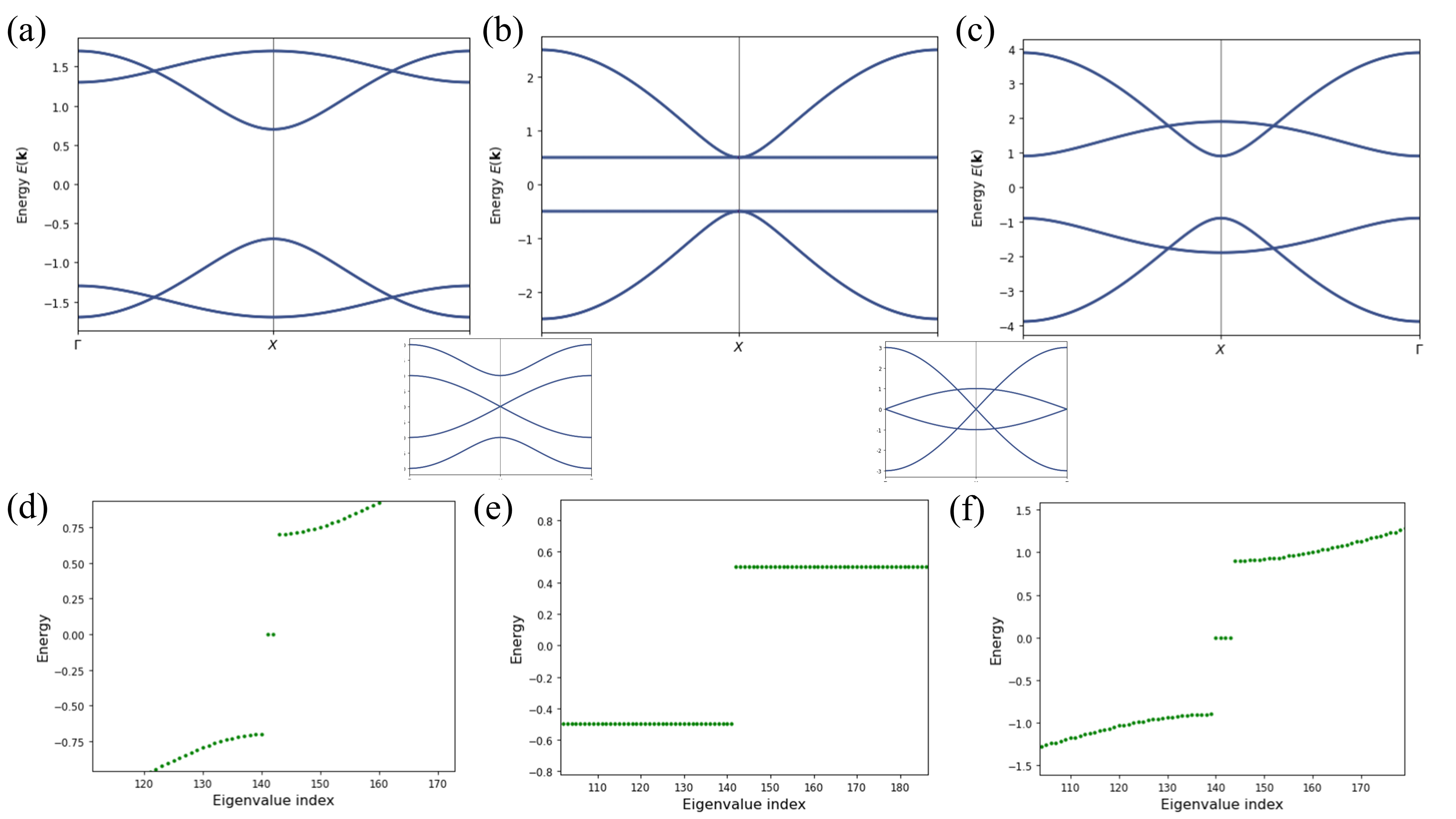} 
    \centering
    \caption{(a) bulk band structure for a parameter set with $t_h=1.0$, $t'_h=0.5$, $t_d=0.5$ and 
    $t'_d=-0.7$, which corresponds to a $\nu=1$ phase.
    (b) bulk band structure for a parameter set with $t_h=1.0$, $t'_h=0.5$, $t_d=0.5$ and 
    $t'_d=0.5$, which corresponds to a $\nu=0$ phase.
    (c) bulk band structure for a parameter set with $t_h=1.0$, $t'_h=0.5$, $t_d=0.5$ and 
    $t'_d=1.5$, which corresponds to a $\nu=2$ phase. Panel (d)--(f) represent the energy spectrum of a finite structure at representative points of the three phases, selected according to the phase-diagram criteria discussed in Sec. \ref{model}.
\label{bulk_bands_edge_spect}}
\end{figure*}
Since two constraints are imposed simultaneously, these regions form codimension-2 sets in the full parameter space. Consequently, they project as isolated points in the $(t_d,t'_d)$ plane. We classify these codimension-2 regions into two groups: the $(\gamma,\gamma')$ pair, which corresponds to genuine topological transition points, and the $(\delta,\delta')$ pair, which corresponds to gap-closing points that are not critical in the topological sense. The latter do not induce a phase transition when crossed along a straight path in the phase diagram. In this sense, we refer to the $(\gamma,\gamma')$ pair as the set of multicritical points, which become planes in the full $(t_h,t'_h,t_d,t'_d)$ parameter space. This completes the classification of the topological transition regions of the CSSH ladder.

The variation of the remaining parameters, $t_{h}$ and $t'_{h}$, will only
changes the form of the phase diagram, but not the general behavior of the topology in the CSSH ladder. In particular, if we fix only $ t_{h}$ for
instance, and vary $t'_{h}$ we will get a three-dimensional space that
is an out-of-plane extension of FIG. \ref{bulk_Phase_diag}.
If we further restrict to the regime $t_h>t'_h$, then, as $t'_h$ approaches $t_h$, the central $\nu=0$ region becomes progressively narrower because the lines $L_{+,\pi}$ and $L_{-,\pi}$ move closer together. This region ultimately disappears when $|t_h|=t'_h$.
In this case a gap closing condition is globally
imposed, which means that bands will touch either at $k=0$ or $k=\pi$ invariably. This
is independent of the values of $t_{d}$ and $t'_{d}$. Nevertheless, the band touching
will, in general, not happen at zero energy. To get a zero-energy gap closing at $|t_h|=t'_h$ it is require the additional condition $t_d=t'_d$
This defines codimension-2 regions that are distinct from the codimension-2 regions projected as $\gamma$ and $\gamma'$ in FIG. \ref{bulk_Phase_diag}, since they involve two gap closings occurring at the same momentum point (not shown in the figure). 
Finally, the phase diagram for the $t_{h} < t'_{h}$ regime will presents the three values for
the $\nu $-phases but with a different location of the regions and also of the critical zones.

To illustrate the band behavior associated with the topological transitions, 
we choose a trajectory in the projected phase space in which $t_d$ is kept constant while $t'_d$ is varied, thereby parametrizing a vertical straight line. 
The constant value of $t_d$ is chosen such that this vertical line passes through the $\delta$ point in FIG. \ref{bulk_Phase_diag}. The selected range of $t'_d$ values drives the system from the $\nu=1$ phase, through a region adjacent to 
the $\nu=0$ phase, across the $\delta$ point, and finally into the $\nu=2$ phase.  The resulting evolution is shown in FIG. \ref{bulk_bands_edge_spect}(a)--(c).  The band structure displays the expected signatures of topological transitions, where a single gap closing occurs when going from $\nu=1$ to $\nu=0$, as illustrated 
in the intermediate panel between FIG. \ref{bulk_bands_edge_spect}(a) and 3(b), whereas two gap closings occur when going from $\nu=0$ to $\nu=2$, as shown in the intermediate panel between FIG \ref{bulk_bands_edge_spect}(b) and 3(c).

Once the nontrivial topological phases have been identified within the AZ classification through the corresponding bulk invariant, the next step is to analyze the bulk-boundary correspondence in a finite system. According to this principle, the edge response is directly determined by the value of the bulk invariant, and for the BDI symmetry class it predicts the presence of $2\nu$ zero-energy edge states. To verify this prediction, we calculate the energy spectrum of a finite CSSH ladder at representative points of the three phases identified in the phase diagram. The corresponding spectra, shown in Fig.~\ref{bulk_bands_edge_spect}(d)--(f) along the same trajectory considered above, clearly exhibit zero-energy modes in the nontrivial phases, while no such states are found in the trivial phase. Moreover, the number of zero-energy states agrees with the value predicted by the bulk invariant, confirming the bulk-boundary correspondence. These edge states are protected by the chiral symmetry characteristic of the BDI class.

\section{Transport signatures of topology}

\begin{figure*}[ht]
    \includegraphics[width=1.\linewidth]{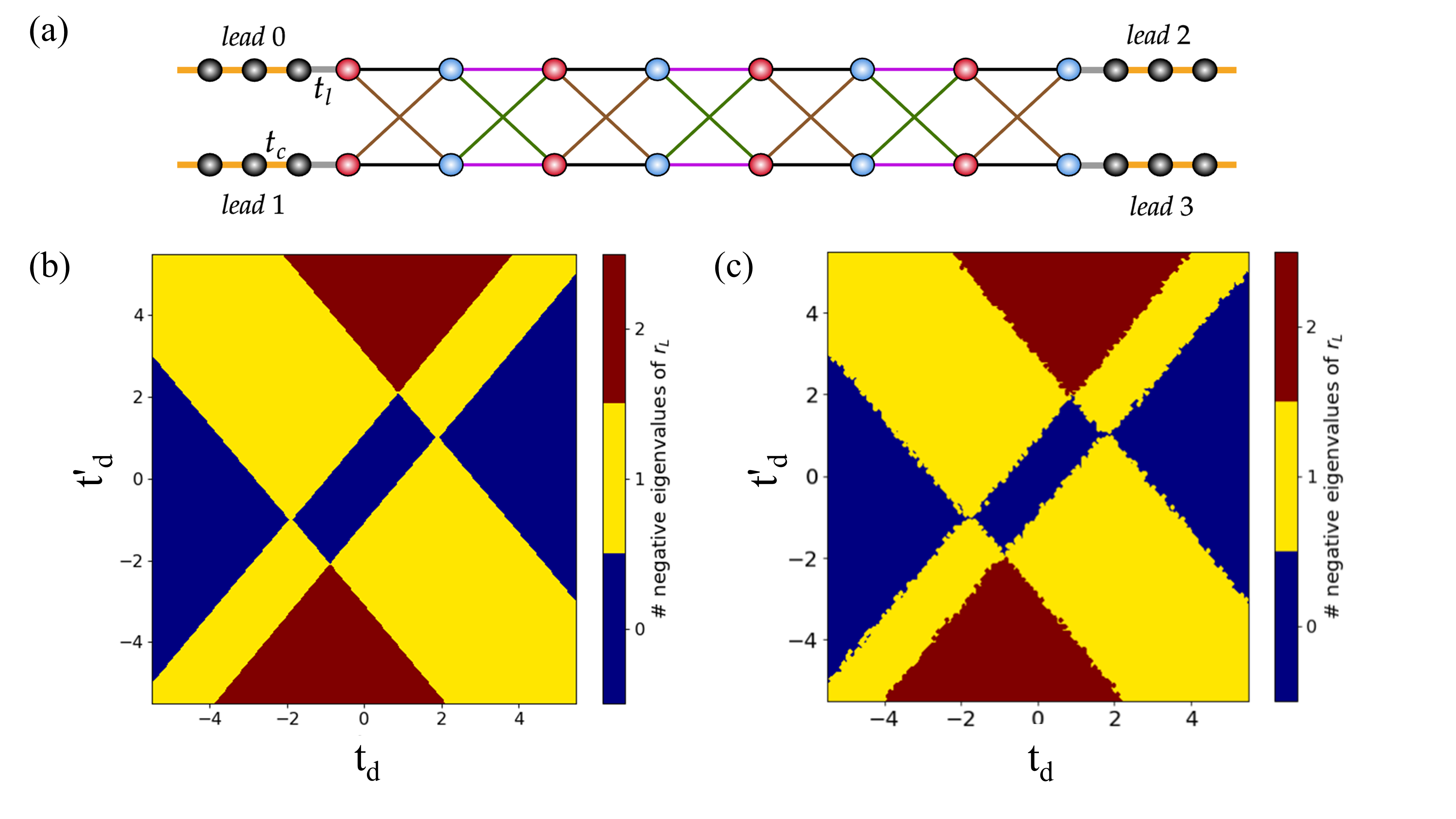}
    \centering
    \caption{(a) Transport configuration used for the four-terminal calculations, including 
the lead labeling.  (b) Signature of the eigenvalues of the reflection matrix $r_L$, corresponding to the topological invariant $\nu$, in the projected $(t_d,t'_d)$ plane for a 
pristine system, where all hoppings of the same type have equal magnitude. The fixed parameters are $t_h=1.5$ and $t'_h=1$. 
(c) Signature of the eigenvalues of the reflection matrix $r_L$, corresponding to the topological invariant $\nu$, in the projected $(t_d,t'_d)$ plane for a  system with random $t_h$ hoppings uniformly distributed in the range $1.5\pm0.2$, with fixed $t'_h=1$. For both calculations in (b) and (c), the 
parameter-space grid has size $201\times201$.}
    \label{long_chain_rL_eig}
\end{figure*}

In addition to the standard procedure for analyzing the topology of the CSSH ladder, in this section we introduce a second characterization focused on transport properties of the system. We consider a finite chain regime, where the localization length of the edge states is negligible, and the system exhibits robust topological properties associated with the bulk invariant. Thus, a four-terminal configuration is employed, in which simple one-dimensional chains act as leads and are connected to each of the four edge sites of the finite CSSH ladder. A graphical representation of the setup is shown in FIG. \ref{long_chain_rL_eig} (a).

Here, we study the quantum transport properties by using the scattering-matrix ($S$-matrix) approach \cite{datta_1995,Lesovik_2011}, with numerical calculations performed using the KWANT package \cite{KWANT_2014}. For the transport configuration considered here, in which incoming states are injected from the left lead, the $S$ matrix is a $4\times4$ matrix. Following the channel labeling shown in FIG. \ref{long_chain_rL_eig}(a), it can be written as:
\begin{equation}
S\ =\ \begin{pmatrix}
r_{00} & r_{01} & t_{02} & t_{03}\\
r_{10} & r_{11} & t_{12} & t_{13}\\
t_{20} & t_{21} & r_{22} & r_{23}\\
t_{30} & t_{31} & r_{32} & r_{33}
\end{pmatrix} =\begin{pmatrix}
r_{L} & t_{LR}\\
t_{RL} & r_{R}
\end{pmatrix}.
\end{equation}

Here, the $2\times2$ matrix blocks $r_{n}$ and $t_{n,m}$, with $n,m\in \{L,R\}$, represent the reflection and transmission  coefficient for left ($L$) and right ($R$) edges of the ladder.
 
A significant limitation arises when the system is characterized using standard channel-resolved reflection or transmission probabilities, defined as $R_{nm}=r_n^{\dagger}r_m$ and $T_{nm}=t_n^{\dagger}t_m$, respectively. This approach provides only a partial resolution of the topological phases. In the near zero energy regime, where the topological information must be inferred from the real-space scattering response, these reflection probabilities do not retain the full state information and reveal only the parity of the topological phase. Therefore, extracting the complete topological information from transport requires access to the full reflection block $r_{n}$ of the $S$-matrix, rather than only to channel-resolved probabilities. As shown in Ref. \cite{PRB_fulga_2012}, for systems in the BDI class, the topological invariant $\nu$ is given by the number of negative eigenvalues of the reflection matrix at zero energy. Following this scheme, we compute the signature of the zero-energy eigenvalues of the reflection block $r_{n}$. Consistent with the discussion above, we restrict the analysis to the projected $(t_d,t'_d)$ plane introduced in Sec. \ref{model}. The resulting signature map, shown in FIG. \ref{long_chain_rL_eig}(b), closely reproduces the analytical phase diagram presented in 
FIG. \ref{bulk_Phase_diag}. The main challenge lies in resolving the 
$(\delta,\delta',\gamma,\gamma')$ points, where the convergence of two or more distinct phases can give rise to minor numerical instabilities. 
Nevertheless, this transport representation of the reflection invariant offers several advantages, as it is defined in real space for finite systems and can be computed efficiently. This approach can be readily generalized to other geometries that preserve the BDI class. A key consideration is the ability to probe the robustness of the reflection invariant when disorder is introduced into certain hopping terms. Disorder is introduced into either the $t_h$ or $t'_h$ parameter, and a grid of values in the ($t_d, t'_d$) plane is explored. Since the phase diagram yields an integer invariant for each pair of points, the resulting values are compared for a random sample of $t_h$ values, with $t'_h$ held fixed. 
A range of values within the interval $\Delta t'_h<1$ is selected for the random sample using a uniform distribution. The center of this interval serves as the reference point for comparison with a pristine system.
An illustrative calculation is presented in the projected phase diagram shown in FIG. \ref{long_chain_rL_eig}.(c) for low disorder, defined as $\Delta t'_h \approx 0.1$ (see the figure caption for the used parameters). In this regime, all analyzed outcomes display consistent trends. Specifically, the general structure of the phase diagram, as determined by the value of the invariant in regions distant from the transition lines or points, remains unchanged. The influence of disorder is most evident in the phase transition regions, where the sharpness of the critical boundaries diminishes. This effect is particularly evident near the codimension-2 regions, which correspond to singular points in the analytical description but become diffuse regions that are no longer sharply defined in the numerical calculations. Nevertheless, the results provide numerical evidence that the topological phases remain robust in the presence of disorder, which is effectively described by the reflection invariant, provided that the global symmetries of the BDI class are preserved.

\section{Flatbands and BICs} 

The analysis above provides a complete characterization of the possible topological 
critical regions of the CSSH model. The only regime discussed separately is the emergence of flat bands for specific values of the hopping parameters, since this 
phenomenon is independent of the topological classification. The conditions for flat band formation in this ladder can then be clarified by analyzing the dispersion relation $E(k)$, which is concisely summarized for the four bands of the model as follows:
\begin{equation}
\begin{aligned}
    E_{\pm,\pm}(k) =& \pm|F_{\pm}(k)|\\ 
    =& \pm \sqrt{A_{\pm}^2+B_{\pm}^2+A_{\pm}B_{\pm}\cos(k)}.
\end{aligned}
\label{dispersions}
\end{equation}

Therefore, a necessary condition is that $E(k)$ remains independent of momentum, which requires the last term inside the square root in equation~(\ref{dispersions}) to be zero (where $A_{\pm}=t_h\pm t_d$ and $B_{\pm}=t'_h\pm t'_d$). If both conditions are satisfied simultaneously, i.e., $A_{\pm}=B_{\pm}=0$, a single flat band at energy $E=0$ emerges in the system. On the other hand, if only one of these conditions is satisfied, either $A_{\pm}=0$ or $B_{\pm}=0$, a codimension-1 region emerges. Since momentum does not influence these bands, the codimension-1 regions are analyzed in $t$-space, as shown in the projected diagram in FIG. \ref{bulk_Phase_diag}. In this representation, flat bands appear as lines parallel to one of the ($t'_h, t'_d$) axes. The intersections of these lines define codimension-2 zones, which are projected onto the same phase diagram as points. Mathematically, these points correspond to parameter values where both zero conditions are satisfied. Notably, two of these intersection points ($\delta$ and $\delta'$) coincide with previously identified non-critical points, while the other two, labeled $\eta$ and $\eta'$, are not frontier points and can also be considered non-critical. Regarding band degeneracy, in a flat band region away from codimension-2 points, the spectrum contains two nondegenerate flat bands, as shown in FIG. \ref{bulk_bands_edge_spect}(b). At the intersection points, these bands merge into a twofold degenerate flat band.

\begin{figure}[ht]
    \includegraphics[width=\linewidth]{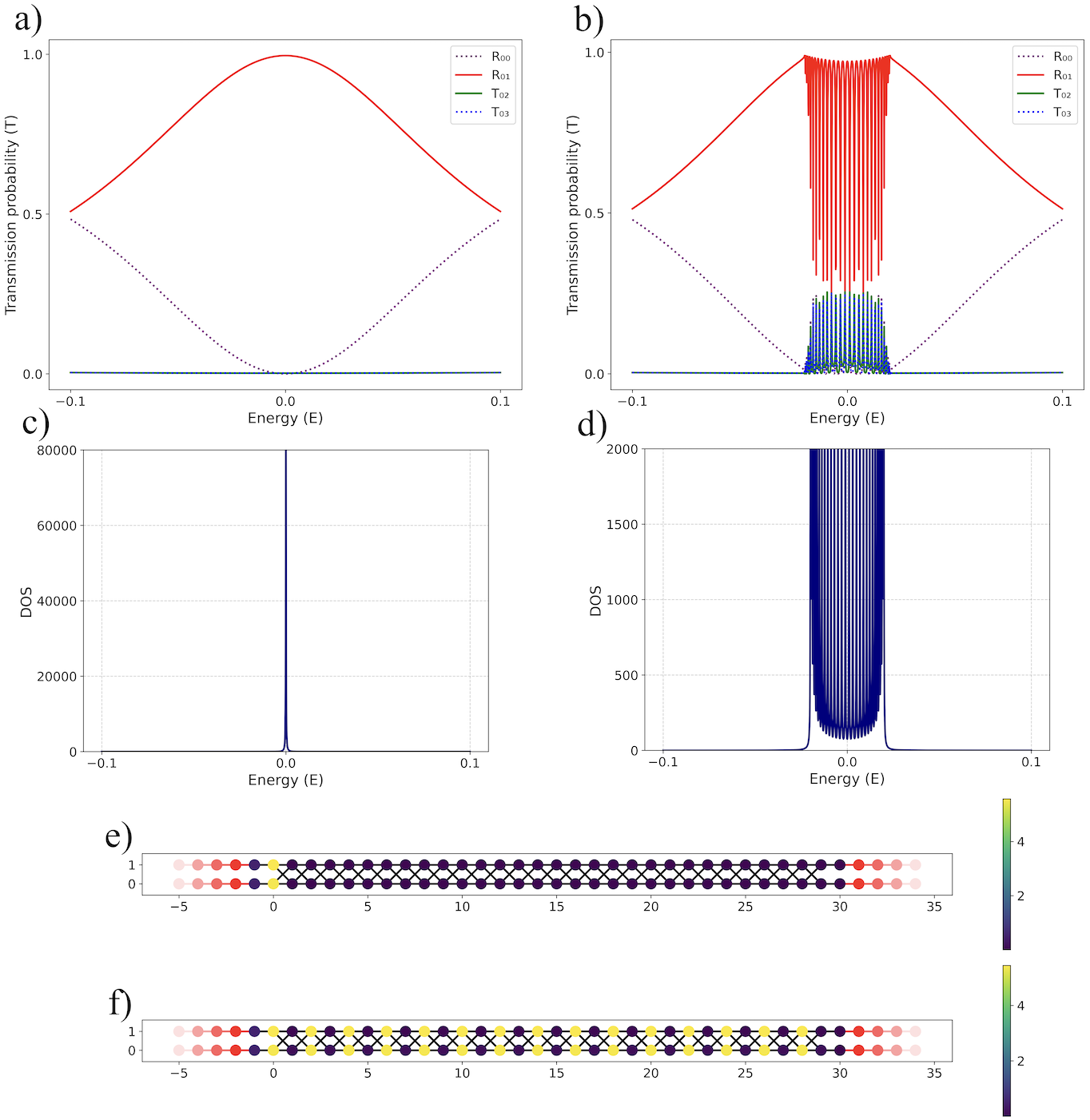}
\caption{(a) Transmission probability in the flat band condition for $t_h=t_h'=t_d=t_d'=1$. (b) Transmission probability slightly off the flat band condition for $t_h=t_h'=1$, $t_d=1.01$, $t_d'=1.01$. (c,d) Density of states for (c) the flat band condition and (d) slightly off the flat band condition. (e,f) Local density of states where the in-signal enters through the lead 0 for (e) the flat band condition and (f) slightly off the flat band condition} \label{BP}
\end{figure}

\begin{figure}[ht]
    \includegraphics[width=.975\linewidth]{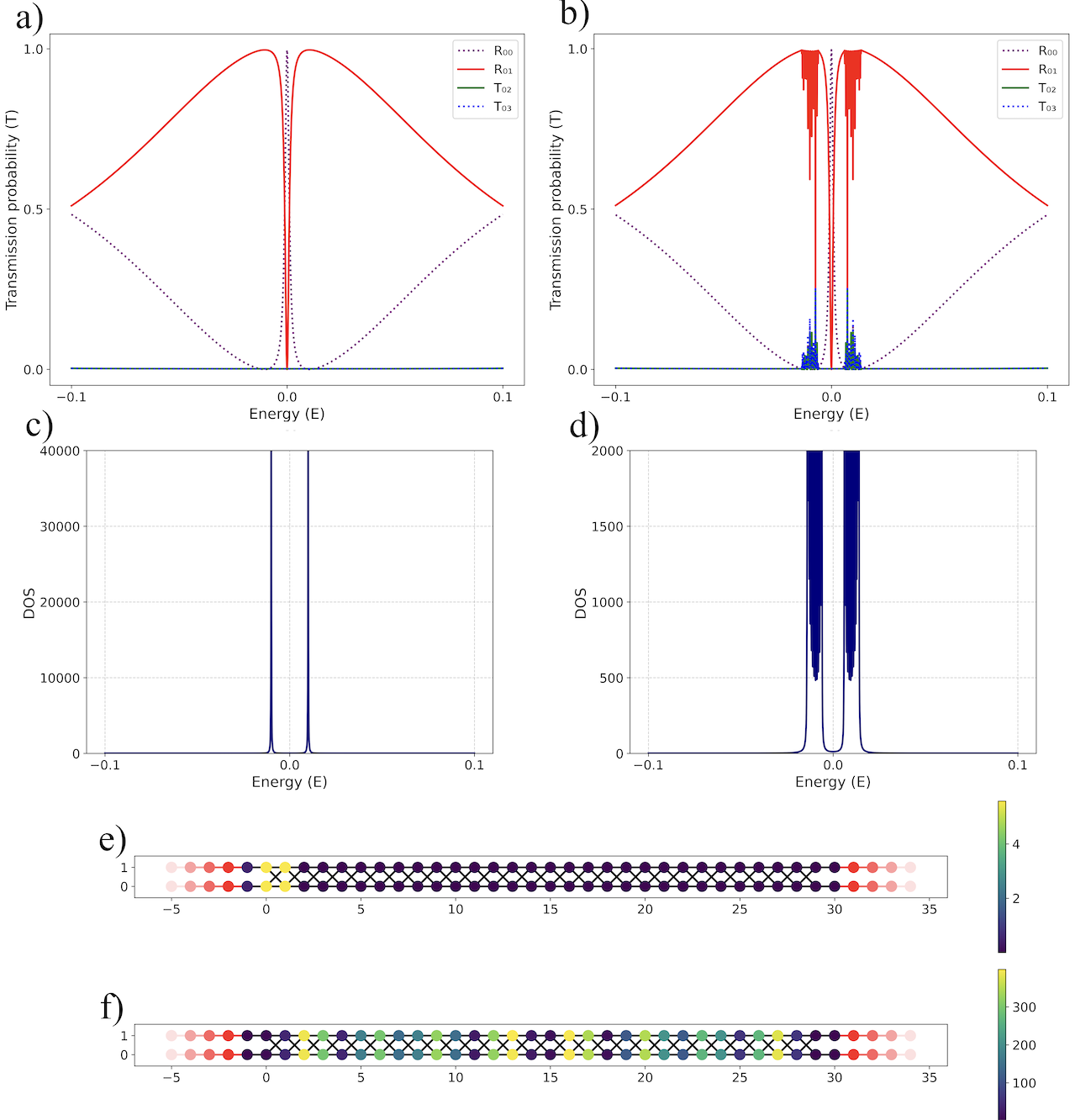}
\caption{(a) Transmission probability in the flat band condition for $t_h=t_h'=t_d'=1$, $t_d=1.01$. (b) Transmission probability slightly off the flat band condition for $t_h=t_h'=1$, $t_d=1.01$, $t_d'=1.004$. (c,d) Density of states for (c) the flat band condition and (d) slightly off the flat band condition. (e,f) Local density of states where the in-signal enters through the lead 0 at $E=0.01$ for (e) the flat band condition and (f) slightly off the flat band condition} \label{BP2}
\end{figure}

When the flat-band condition is satisfied, the effective antisymmetric chain becomes completely decoupled from the continuum, giving rise to an $N$-fold degenerate BIC at $E=0$, where $N$ is the number of sites in this chain. For $t_h=t_d=1$ and $t'_h=t'_d=1$, this BIC appears as a single $\delta$-like peak in the DOS of Fig.~\ref{BP}(c). Since it is completely decoupled from the continuum, it has no direct projection on the transmission shown in Fig.~\ref{BP}(a). Nevertheless, the system presents an almost complete transmission through lead 1 at zero energy, similar to the behavior of the topological phase with $\nu=1$. This transport feature is produced by the states coupled to the leads and not by the BIC itself. Therefore, although the system remains in the trivial phase with $\nu=0$, its zero-energy response resembles that of a system with a pair of boundary states. This behavior can also be seen in the LDOS of Fig.~\ref{BP}(e), calculated for an incoming signal through lead 0, where the spectral weight is mainly localized at the left boundary.

When a small asymmetry is introduced by setting $t_d=1.01$ and $t'_d=1.01$, the symmetry protecting the BIC is broken. Consequently, the $N$-fold degeneracy is lifted, and the original BIC splits into $N$ quasi-BICs associated with the states of the effective antisymmetric chain. These states become weakly coupled to the continuum, and the single peak at $E=0$ evolves into $N$ $\delta$-like peaks distributed within the interval $-0.01\lesssim E\lesssim0.01$, as shown in Fig.~\ref{BP}(d). This weak coupling also gives the states a finite projection on the transmission, producing the multiple antiresonances observed in Fig.~\ref{BP}(b). The corresponding LDOS in Fig.~\ref{BP}(f) shows that the quasi-BICs are no longer restricted to the boundary but extend through the bulk and acquire a small contribution in the leads. However, their spatial distribution still follows a pattern related to the original symmetry, indicating that the antisymmetric chain remains only weakly coupled to the continuum.

When only one of the conditions $A_{\pm}=0$ or $B_{\pm}=0$ is satisfied, two flat bands emerge at finite energies symmetrically located around the Fermi energy. This case is shown in Fig.~\ref{BP2} by setting $t_d=1.01$, which places the flat bands at approximately $E=\pm0.01$. In the DOS of Fig.~\ref{BP2}(c), they appear as two $\delta$-like peaks, each one associated with an $N$-fold degenerate set of BICs, where $N$ is the number of sites in the effective antisymmetric chain. Because these states are completely decoupled from the continuum, they have no direct projection on the transmission shown in Fig.~\ref{BP2}(a). In addition, the absence of states at $E=0$ produces an antiresonance around the Fermi energy. The LDOS of the BIC at $E=0.01$ is shown in Fig.~\ref{BP2}(e), where the spectral weight is mainly localized at the left boundary.

When an additional asymmetry is introduced, by setting $t'_d=1.004$, the symmetry protecting the BICs is broken and the degeneracy of each set is lifted. Consequently, the two $\delta$-like peaks split into two collections of quasi-BICs distributed near $E=\pm0.01$, as shown in the DOS of Fig.~\ref{BP2}(d). Since these states are now weakly coupled to the continuum, they acquire a finite projection on the transmission and produce the multiple antiresonances observed in Fig.~\ref{BP2}(b). This coupling is also visible in the LDOS at $E=0.01$ shown in Fig.~\ref{BP2}(f), where the spectral weight extends from the boundary into the bulk and acquires a small contribution in the leads. This distribution confirms that the quasi-BICs interact with the continuum while still preserving part of their localized character.

\section{Final Remarks}

In this work, we studied the electronic, topological, and transport properties of a Creutz-Su-Schrieffer-Heeger ladder in the absence of a magnetic field. By decomposing the system into two effective SSH chains, we obtained a clear interpretation of its bulk topology and showed that the total winding number is given by the sum of the invariants of the two independent channels. This approach allowed us to identify the trivial phase with $\nu=0$ and the nontrivial phases with $\nu=1$ and $\nu=2$, as well as the codimension-1 and codimension-2 regions separating them. The energy spectrum of a finite ladder confirms the bulk-boundary correspondence, with the number of zero-energy edge states determined by the value of the winding number.

We also showed that the topology of the ladder can be identified from its transport response. In particular, the number of negative eigenvalues of the zero-energy reflection matrix reproduces the complete topological phase diagram, including phases that cannot be distinguished using only channel-resolved transmission and reflection probabilities. This scattering-matrix invariant remains stable under weak hopping disorder as long as the symmetries of the BDI class are preserved. Thus, transport provides a direct real-space method to characterize the topological phases of finite CSSH ladders without requiring an explicit calculation of the bulk invariant.

Finally, we derived the conditions for the formation of non-topological flat bands and established their direct connection with BICs. At the exact flat-band condition, the corresponding states are decoupled from the continuum and appear as highly degenerate $\delta$-like peaks in the DOS, without a direct projection on the transmission. When the protecting symmetry is weakly broken, their degeneracy is lifted and the BICs evolve into collections of quasi-BICs that couple weakly to the continuum, producing multiple antiresonances in the transmission. These results show that topological edge states and BICs coexist in the CSSH ladder but originate from different localization mechanisms. The possibility of distinguishing and controlling these states through transport makes this system a promising platform for studying the interplay between topology, destructive interference, and localization in quantum systems.

\section*{Acknowledgements}
This work was partially financed by Fondecyt grant 1220700. K.A.G. acknowledges an ANID doctoral fellowship. S.B. acknowledges a DGIIE-USM postdoctoral fellowship.

\vspace{1cm}

\bibliographystyle{elsarticle-num}
\bibliography{references}

\end{document}